\documentclass[aps,pre,nofootinbib,showpacs,twocolumn]{revtex4}%
\usepackage{amssymb,latexsym,mathrsfs}
\usepackage{amsfonts}
\usepackage{graphicx,graphics,epsf}

\newcommand{\be}{\begin{equation}}
\newcommand{\ee}{\end{equation}}
\newcommand{\bea}{\begin{eqnarray}}
\newcommand{\eea}{\end{eqnarray}}
 
\begin{document}
 
\title{Modeling wealth distribution in  growing markets}
\author{Urna Basu }\email{urna.basu@saha.ac.in}
\author{ P. K. Mohanty} 
\affiliation{Saha Institute of Nuclear Physics,1/AF,
Bidhan Nagar, Kolkata 700064, India}

\begin{abstract}
We introduce an auto-regressive model which captures the 
growing nature of realistic markets.  
In our  model agents do not trade with other agents, they interact 
indirectly only through a market. 
Change of their wealth depends, linearly on  how much they invest, 
and stochastically on how much they  gain from the noisy market. 
The average wealth of the market could be fixed or growing. We 
show  that in a market where investment capacity of agents differ, 
average wealth of agents generically follow the Pareto-law. In 
few cases, the individual distribution of wealth  of every agent
could also be obtained exactly. We also show that the underlying  
dynamics of  other well studied kinetic models of markets can be 
mapped to the dynamics of our auto-regressive model.
\end{abstract}

%\begin{document}
\maketitle
 
Kinetic models have been drawing substantial attention as  model 
markets. In these models\cite{Kinetic},  markets are compared 
with systems of ideal gases, where agents and their wealth are 
considered 
analogous to the gas particles and their energy respectively. 
Trading between any pair of agent is  similar to a collision 
process where energy  or wealth is shared between agents. 
Several models of both conserved \cite{Angle,Redner,Yako00,CC,CCM} 
and open \cite{Redner,Mezard,Slanina,Matteo} economic systems have been 
studied recently, which differ mainly  
in their exchange rules,  namely whether collision is elastic or 
inelastic, if a fraction or  the whole energy of a pair is 
shared between agents, etc. A minimal model of a closed market is
when  a randomly chosen pair of particles(agents)  
collide (trade)  elastically such that the total energy(wealth) 
of the pair is shared randomly between the particles (agents).
This wealth conserving dynamics naturally predicts \cite{Yako00} a  
Gibb's distribution of wealth $P(x) \sim exp(-\beta x)$ 
in equilibrium,  which has been observed  in distribution of 
income-tax return of individuals in several 
countries\cite{Silv05}.  However, the tails of the wealth 
distribution in \cite{Silv05} and other economic 
systems\cite{Real} follow  a power-law distribution, as 
predicted by V. Pareto\cite{Pareto} 
(known as {\em Pareto-law}).
In an attempt to get a scale free distribution 
within the frame-work of wealth conserving dynamics, a simple 
model was proposed  by Chatterjee Chakrabarti and Manna 
(CCM)\cite{CCM}, where agents contribute only a fraction of 
their wealth  for trading depending on their savings 
propensities which differ among agents. 
Numerical simulations of this model strongly suggest that
the the distribution of {\it average} wealth 
$\{w_i= \langle x_i\rangle\}$ follow $P(w) \sim w^{-\gamma}$, 
with $\gamma=2$. Later, exact results\cite{PK} show that the  
tail of the distribution is generically scale-free with 
$\gamma=2$. For certain typical variations of the model
one may get $\gamma\ne 2$. Note that in these models, it 
is {\it only} the average wealth which show a scale free 
distribution.  
Wealth of any individual agent is distributed about 
his average following a Gamma{\it -like}\cite{Manna} 
distribution.% (the exact distribution is still an open question). 

Although,  kinetic models \cite{Kinetic} are  successful 
in describing basic features of wealth distribution, they do not 
capture the  growing features  of  most realistic markets. 
Recently, growing markets are modeled by  pouring an extra 
amount of wealth during each trading which is proportional to 
the wealth of one\cite{Redner} or both\cite{Slanina} agents 
participating in trading. A power-law distribution for 
rich was observed only in
\cite{Slanina} where $P(w) \sim w^{-1.7}$. To have 
finite average wealth, the tail of this distribution can not 
be scale-free; it must be cut off at some finite $w$.  
 
In this article, we introduce a  minimal model of  
growing markets and show that this class of models generically 
produce a power-law tail in the  wealth distribution, 
independent of the details of the market and the trading rules. 
Both static and growing markets having conserving 
or non-conserving 
dynamics lead to Pareto-distribution of wealth, 
$P(w)~w^{-\gamma}$ with $\gamma\ge 2$. 
Kinetic models  are just a  sub-class satisfying conservation 
of wealth and their dynamics could be  mapped to the dynamics of 
our model. This exact mapping suggests that wealth conservation is not 
necessary for  the description of markets. It also
provides  a route to capture the exact distribution of the 
fluctuations of wealth of individual agents. Finally, these models, 
being auto-regressive(AR), bridge a connection  between  kinetic models  
studied recently in econophysics and other AR 
models\cite{AR} of markets studied by economists.

First, the model.   Let us take a system of $N$ independent agents 
$i=1\dots N$, whose wealth  at a given time $t$ is  $x_i(t)$. 
Each agent $i$, depending on his  investment capacity $0<\mu_i\le 1$, 
invests a definite fraction of wealth $\mu_i x_i(t)$ in 
the market. The market stochastically returns a  net gain 
$(t)$. Thus, wealth  of agent $i$ at time $t$ is  
\be
 x_i(t) =  (1-\mu_i)x_i(t-1) + \xi(t).  
\label{eq:model}
\ee

In this minimal model $\xi(t)$ is taken as a uncorrelated positive 
stochastic variable  with probability distribution function (PDF) 
$h(\xi)$; it does not depend on  $\{x_i\}$. Thus, agents may gain 
or lose from the market. The auto-regressive nature of the model 
that $x(t)$ depends on $x(t-1)$ is clear from (\ref{eq:model}).

First we will calculate the steady state of 
(\ref{eq:model}). Let us define an operator 
${\cal B}$ which  moves the variables one step backward 
in time, i.e., $x(t-1) = {\cal B} x(t)$. For 
convenience let us take $\lambda_i\equiv 1-\mu_i$, 
which is similar to the savings 
propensity  defined in  kinetic models \cite{Kinetic}. 
Now, (\ref{eq:model})  can be written as, 
\bea
x(t)&=& \frac{1}{1-\lambda {\cal B} }  \xi(t)=\sum_{n=0}^{\infty} \lambda^n \xi(t-n)\cr
&=&\sum_{n=0}^{t} \lambda^n \xi(n),
\label{eq:weightedsum}
\eea
  where we have dropped the index $i$ as agents are independent. In  
the last step  we have used the fact that $\xi(t)$ is a
uncorrelated random variable and that $\xi(n<0) =0$. 
% We have also assumed an  initial condition that $x(0)= \xi(0)$.
Thus, the steady state distribution   $P(x)$
which is reached  as  $t\to \infty$ is the PDF of the 
stochastic variable 
\be
x=  \sum_{n=0}^{\infty} \lambda^n \xi(n)
\label{eq:ss}
\ee
which is  just  a weighted sum  of $\{ \xi(n)\}$ with weights$\{\lambda^n\}$.
Let $x_m =\sum\limits_{n=0}^{m} \lambda^n \xi(n)$ be  the first $m$ terms of  
(\ref{eq:ss}) and their distribution be $P_m(x)$.  
From (\ref{eq:weightedsum}) and (\ref{eq:ss}) it is clear that 
$x_m = x(t=m)$. It implies, first that  true steady  state  gets contributions 
from all orders of $\lambda^n$. Secondly, $P_m(x)$ can be considered  as the 
distribution at $t=m$.   

Since $x_m = \lambda^m \xi(m) +x_{m-1}$, 
$P_m(x)$ satisfies  a recursion relation, 
\be
P_m(x) =\frac{1}{\lambda^m} \int_0^x P_{m-1}(y) h(\frac{x-y}{\lambda^m})dy.
\label{eq:int}
\ee 
The steady state distribution is then  $P(x)\equiv \lim \limits_{m\to\infty} P_m(x).$
Clearly, from  (\ref{eq:int})  one can see that 
\be 
P_m(0) = 0 ~~ \textrm{for~ all} ~ ~m>0. 
\label{eq:bc}
\ee
Thus in steady state we must have $P(x=0)=0$. 
Equation (\ref{eq:bc}),  being independent  of the choice of 
$h(\xi)$,  can be used as generic boundary conditions for (\ref{eq:int}). 
Secondly,  it indicates that  the steady state distribution 
is neither Gibb's nor Pareto like, where  $P(x=0)$ is finite.  

To proceed further, we need to be more specific, 
namely  we need to know  $h(\xi)$. Before  considering the 
generic growing markets, we   
consider few examples of  {\it static} markets where 
average wealth of the market $a\equiv \langle \xi \rangle$ 
is fixed.
 
\begin{itemize}
\item{\it Normal distribution of $\xi$ :} The first example is when fluctuation
of the market is normal, $i.e$,  $h(\xi)$  is a Gaussian distribution 
denoted by ${\cal G}(\alpha_0, \sigma_0)$ with mean  $\alpha_0$  and 
standard deviation $\sigma_0$. In this case, 
the steady state distribution  $P(x)$ is ${\cal G}(\alpha, \sigma)$ where
\be
\alpha = \frac{\alpha_0}{1-\lambda}~~ {\rm and } ~~  
\sigma= \frac{\sigma_0}
{\sqrt{1-\lambda^2}}.
\label{eq:normal}
\ee
It is easy to check that  ${\cal G}(\alpha, \sigma)$  satisfy 
Eq. (\ref{eq:model}) in steady state; i.e, if PDF of $x$ and $\xi$ are  
${\cal G}(\mu_0, \sigma_0)$ and   
${\cal G}(\mu, \sigma)$ respectively, then PDF of $\lambda x+\xi$ 
is same as PDF of $x$. Note that agents in this case can  have  
negative wealth even though $\langle x\rangle > 0$. The 
negative wealth may be interpreted as {\em debt}.

\item{\it Exponential distribution of $\xi$:}  
In the next example we take $h(\xi)=\exp(-\xi)$. 
This case is interesting, because for $\lambda=0$  it gives 
same steady state distribution as that of the CC model\cite{CC}, 
$i.e$ $P(x) = \exp(-x)$.  For non zero $\lambda$,  
we need   to solve the integral equation  (\ref{eq:int}).  
Instead we rewrite it 
as a differential equation (which is possible in this case),
\be
\frac{d}{dx}P_m(x)  = \frac{1}{\lambda^m}  \left[  P_{m-1}(x) -P_m(x) \right],
\label{eq:diff}
\ee
where $m>0$, and the  boundary conditions  are given by Eq. (\ref{eq:bc}). 
For $m=0$, $P_0(x)\equiv h(x)$. 
In terms of $G_m(s)$,  which is the Laplace transform (LT) of $P_m(x)$, 
Eq. (\ref{eq:diff})  becomes a difference equation 
\be
G_m(s) = \frac{1}{1+\lambda^m s} G_{m-1} (s),
\label{eq:LT}
\ee
 whose formal solution is $$G_m(s)=\prod\limits_{k=0}^{m-1} (1+\lambda^k s)^{-1}  G_0(s).$$
 Again, let us  remind that $G_0(s)$ is the LT of $P_0(x)=h(x).$   
Finally, $P(x)$ is the inverse LT of 
\be 
G(s) =  \prod_{k=1}^{\infty}  \frac{1}{1+\lambda^k s}  G_0(s),
\ee
which can be written as the following series :  
\bea
&&P(x) =  \sum_{m=1}^\infty C_m \exp(-x/\lambda^m) \cr 
&& {\rm where ~~} C_m^{-1} = \lambda^m \prod_{0<n\ne m}^{\infty} (1-\lambda^{n-m}).
\label{eq:exact}
\eea
 
Although Eq. (\ref{eq:exact})  is an infinite series, first few terms are 
good enough for numerical evaluation of the distribution.  Terms up 
to $m=n$ gives $P_n(x)$, which can be interpreted either as an approximation of 
true steady state distribution $P(x)$ to $n^{th}$ order in $\lambda$ or 
as the  distribution at finite time  $t=n$.  In Fig. \ref{fig:compare} we 
compare $P(x)$ which  is obtained numerically with  the first 
four terms of (\ref{eq:exact}) for  $\lambda=0.4$.  
Note, that $P(x)$  is a Gamma-like 
distribution similar to what has been obtained in \cite{Redner,CC,Manna}.

\begin{figure}
\vspace*{0.2cm}
\centerline{\includegraphics[width=6cm]{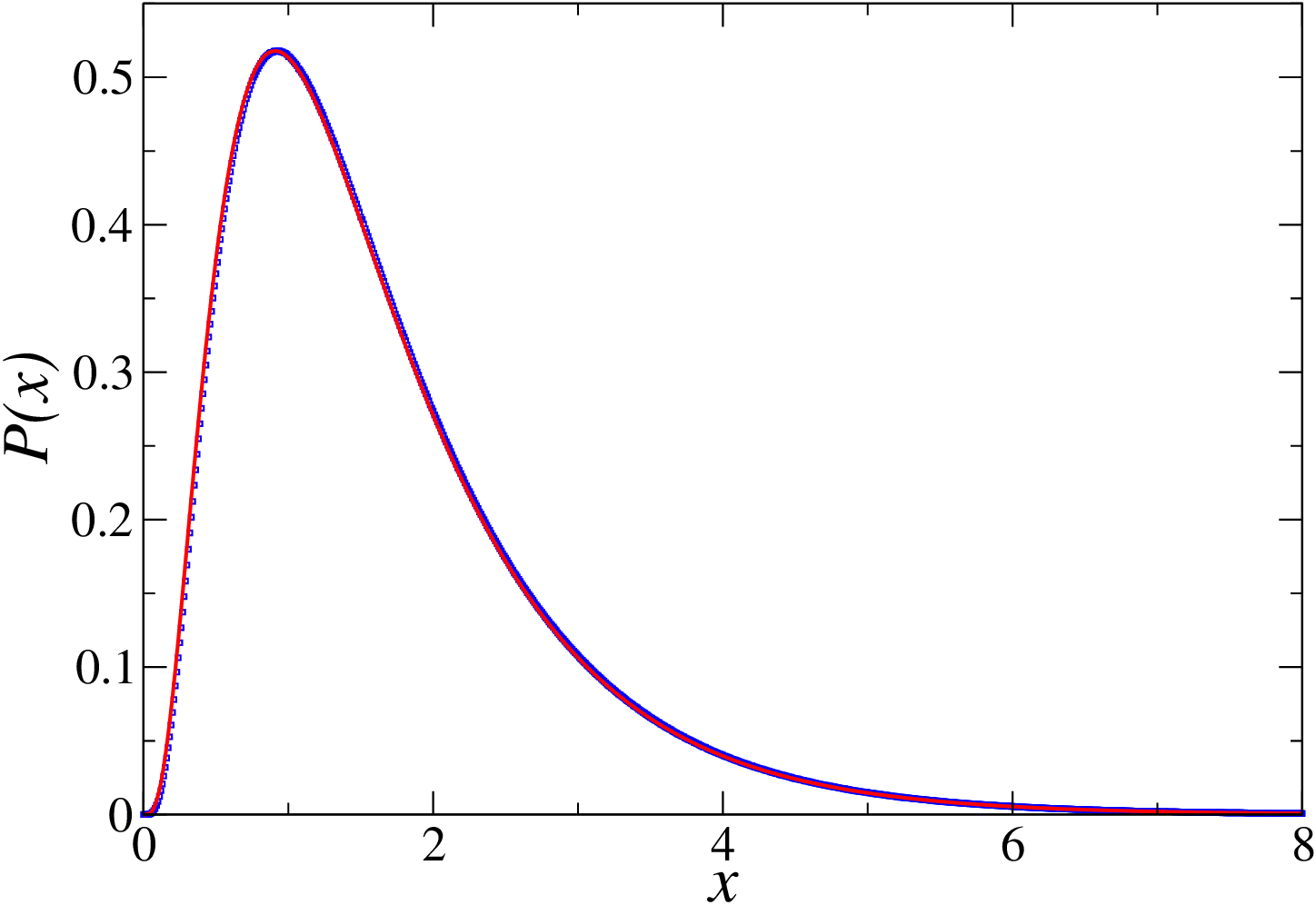}}
\caption{$P(x)$ as a function of $x$, for $h(\xi) = exp(-\xi)$. 
The numerically obtained steady state distribution(for $\lambda=0.4$) 
of wealth (symbols) is compared with the first four terms of 
$P(x)$  from Eq.  (\ref{eq:exact}), drawn as solid line.}  
\label{fig:compare}
\end{figure}

\item{\it Pareto-law :}
In our model, the  wealth distributions  of individual agents are not 
simple and depend  on their investment capacities $\mu_i$. 
Their averages, however, follow a power-law. To prove this let us define 
$\langle x_i \rangle= w_i$. In steady state 
$\langle x(t)\rangle= \langle x(t-1) \rangle$. Thus,  
Eq. (\ref{eq:model}) gives  
\be 
w_i = \frac{\langle \xi\rangle}{\mu_i}. 
\label{eq:wi}
\ee
Agents in this model differ in their investment capacities. 
In a system of $N$ agents the average number
of agents  having investment capacity $\mu$  is $Ng(\mu)$ where 
$g(\mu)$  is the distribution  of $\mu$. Thus, we can write 
$w(\mu) =  \langle \xi \rangle /\mu$. Distribution of $w$ is 
then
\be
P(w) = g(\mu) |\frac{d\mu}{dw}| =  \langle \xi\rangle \frac{g(\langle \xi\rangle/w)}{w^2}.
\label{eq:Pw}
\ee 
A similar argument was used in \cite{PK} for  deriving the wealth 
distribution of CCM model. Although, distribution for the rich 
(large $w$) is generically $P(w) \sim w^{-2}$, 
one can obtain $\gamma >2$  in typical cases. For example, 
if PDF of $\mu$ is $g(\mu)=\mu^{\alpha}/(\alpha-1)$ with  
$0\le \alpha<1$, the  asymptotic distribution of (\ref{eq:Pw}) 
results $P(w) \sim w^{-\gamma}$, where $\gamma= 2+\alpha$.

%%%%%%%%%%%%%%%%%%%%%%%%%%%%%%%%%%%%%%%%%%%%%
% {\bf Negative  $\langle \xi\rangle$}.  
%%%%%%%%%%%%%%%%%%%%%%%%%%%%%%%%%%%%%%%%%%%%%%%

\item{\it Growing markets :}  The kinetic models  of markets 
\cite{Yako00,CC,CCM} are 
defined   with  wealth conserving dynamics, which keeps the total wealth of 
the system  constant. In our model, we can easily incorporate the growth 
feature of the market ( say, stock-markets) by introducing explicit time 
dependence  in the distribution of $\xi$. For example, the mean 
$\langle \xi\rangle \equiv a(t)$ may vary in time. The 
distribution of wealth $P(x,t)$ will then depend explicitly on $t$. 
However, in the adiabatic limit, when  $a(t)$ varies slowly 
(such that $a(t-1) \approx a(t)$), we have $P(x,t-1) \approx P(x,t)$. 
In this limit, thus, $P(x,\tau)$  is identical to the steady 
state distribution of the time-independent model, 
where $\xi$ has an average $\langle \xi\rangle = a(\tau)$.  

For demonstration,  we take $h(\xi)$  to be an exponential distribution 
with varying average $a(t)= t/T$. In other words, $h(\xi,t)= \exp[-x/a(t)]/a(t)$.  
From the numerical simulations we calculate the  distribution $P(x,T)$ at $t=T$ 
for  different values of $T$.  Since $a(T)=1$,  $P(x,T)$ is compared with 
the steady state distribution (\ref{eq:exact}). Figure \ref{fig:grow} 
compares the distributions for $T=20$  and $T=200$, which clearly 
suggests that in the quasi-static limit $T\to \infty$ the instantaneous 
distribution depends only on the instantaneous distribution of $\xi$. 

\begin{figure}
\vspace*{.2 cm}
\centerline{\includegraphics[width=6cm]{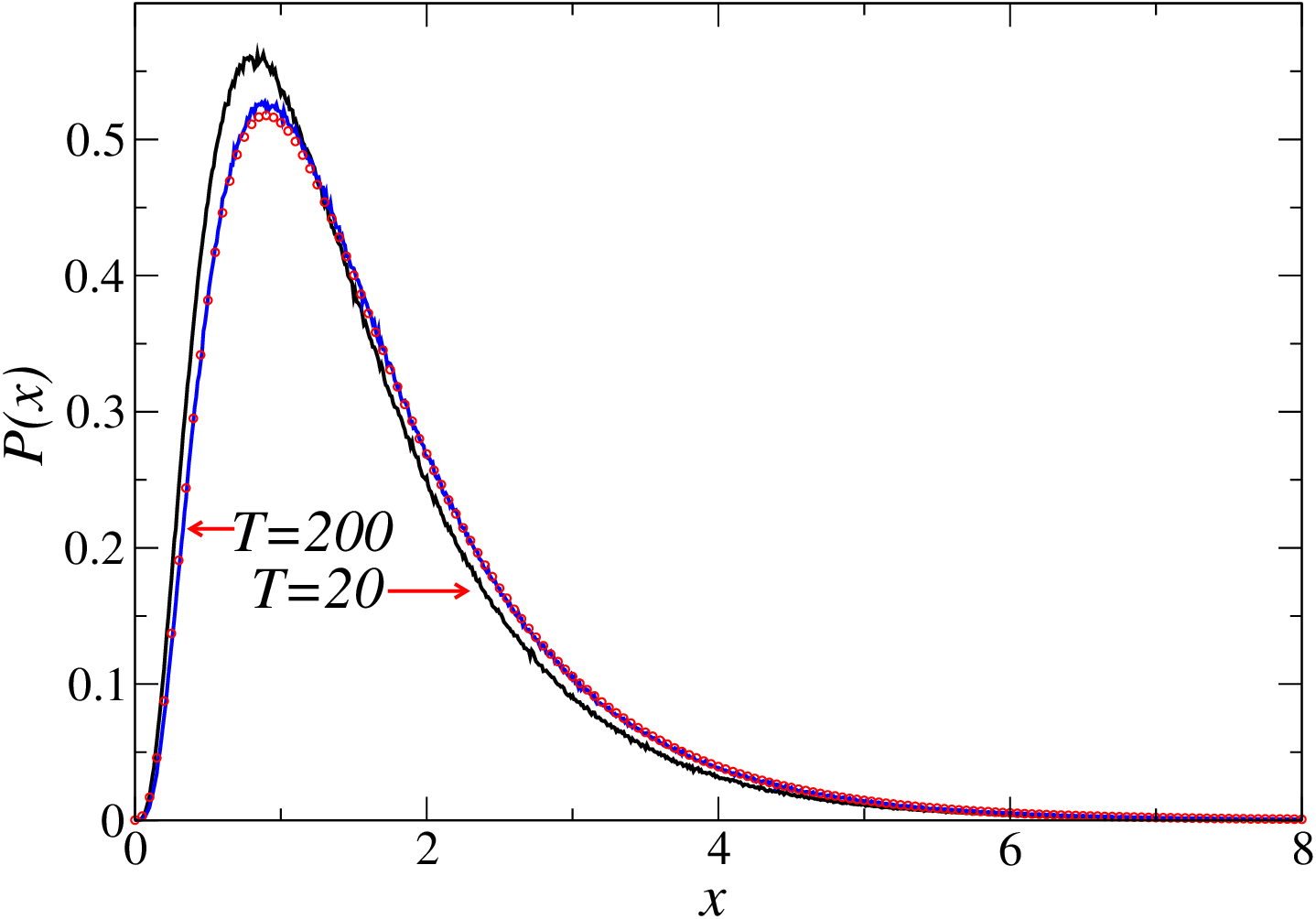}}
\caption{The figure compares the wealth 
distribution $P(x,t=T)$ of growing markets where average wealth $a(t) = t/T$, 
with  that of the {\it static} market with $a(t)=1$ (line). As expected, 
For slowly growing market (large $T$), $P(x,t=T)$ compares well 
with the distribution in $static$ market, suggesting that 'for adiabatically 
growing markets, instantaneous distribution of wealth  is independent of 
the history'.}  
\label{fig:grow}
\end{figure}
 
\item{\it Annealed $\lambda$:}  Another interesting case is when  savings 
propensity of agents change in time. This is modeled by taking $\lambda$ 
as a stochastic variable %. Let's take a specific example, where $\lambda$ is 
distributed, say uniformly in $(0,1)$.  Let $h(\xi)= \exp(-\xi)$. The 
steady state distribution of wealth is then $P(x) = \Gamma_2 (x)= x \exp(-x)$, 
which can be proved as follows.  
If $P(x)= \Gamma_2(x)$ then  $P(u=\lambda x)=exp(-u)$\cite{gamma*uniform}.  
Thus, PDF of  right hand side of Eq. (\ref{eq:model}) is $\Gamma_2(x)$\cite{exp+exp} 
which is same as the PDF of left hand side. This, completes the proof.  

\end{itemize}
 
In rest of the article we  discuss kinetic models studied in the context of 
wealth distribution and show that the dynamics of these models can be 
mapped to  the AR model defined in Eq. (\ref{eq:model}). First 
let us consider the CCM model\cite{CCM}. The main  
idea  in this model is that the  agents, labeled by $i=1\dots N$, are 
considered to have  savings propensities $\{\lambda_i\}$ distributed 
as $g(\lambda)$. During trading, wealth $x_i$ and $x_j$ of two randomly 
chosen agents $i$ and $j$ changes to $x_i^\prime$ and $x_j^\prime$ respectively 
such that 
\bea
x_i^\prime&=& \lambda_i x_i  + r T_{ij}\cr
x_j^\prime&=& \lambda_j x_j + (1-r)T_{ij},
\label{eq:ccm}
\eea
where $T_{ij}= (1-\lambda_i) x_i + (1-\lambda_j) x_j$,  and $r$ is a stochastic 
variable with PDF ${\cal U}(r)$,  uniform in $(0,1)$. 
The  wealth conserving dynamics (\ref{eq:ccm}) can be interpreted as follows. 
Both  agents $(i,j)$ save a fraction ($\lambda_i, \lambda_j$) of 
their wealth and  contribute the rest for trading. The  total trading  
wealth $T_{ij}$ is then randomly shared between 
agents $i$ and $j$. A special case of  the model  
$\{\lambda_i=\lambda\}$ was studied earlier by Chakrabarti and Chakraborti 
(CC)\cite{CC}. Explicitly, the dynamics of the model is
\bea
x_i^\prime&=& \lambda x_i  + \left[ r (1-\lambda) (x_i +x_j)\right]  \cr
x_j^\prime&=& \lambda x_j + \left[ (1-r)(1-\lambda) (x_i +x_j)\right]  ,
\label{eq:cc}
\eea
where a skew wealth distribution was found for  $\lambda\ne 0$.  
Although, the exact steady state wealth distribution $P(x)$  of CC 
model is not known, it can be fitted well to a  Gamma distribution 
$\Gamma_n(x)\equiv  x^{n-1} \exp(-x)/(n-1)!$ with 
$n= (1+2\lambda)/(1-\lambda)$\cite{Gamma}. However, later 
analytical studies \cite{NoGamma} disagree with the 
Gamma distribution. In this article we refer  to it 
as {\it Gamma-like} distribution. 

\begin{figure}
\vspace*{.2 cm}
\centerline{\includegraphics[width=6cm]{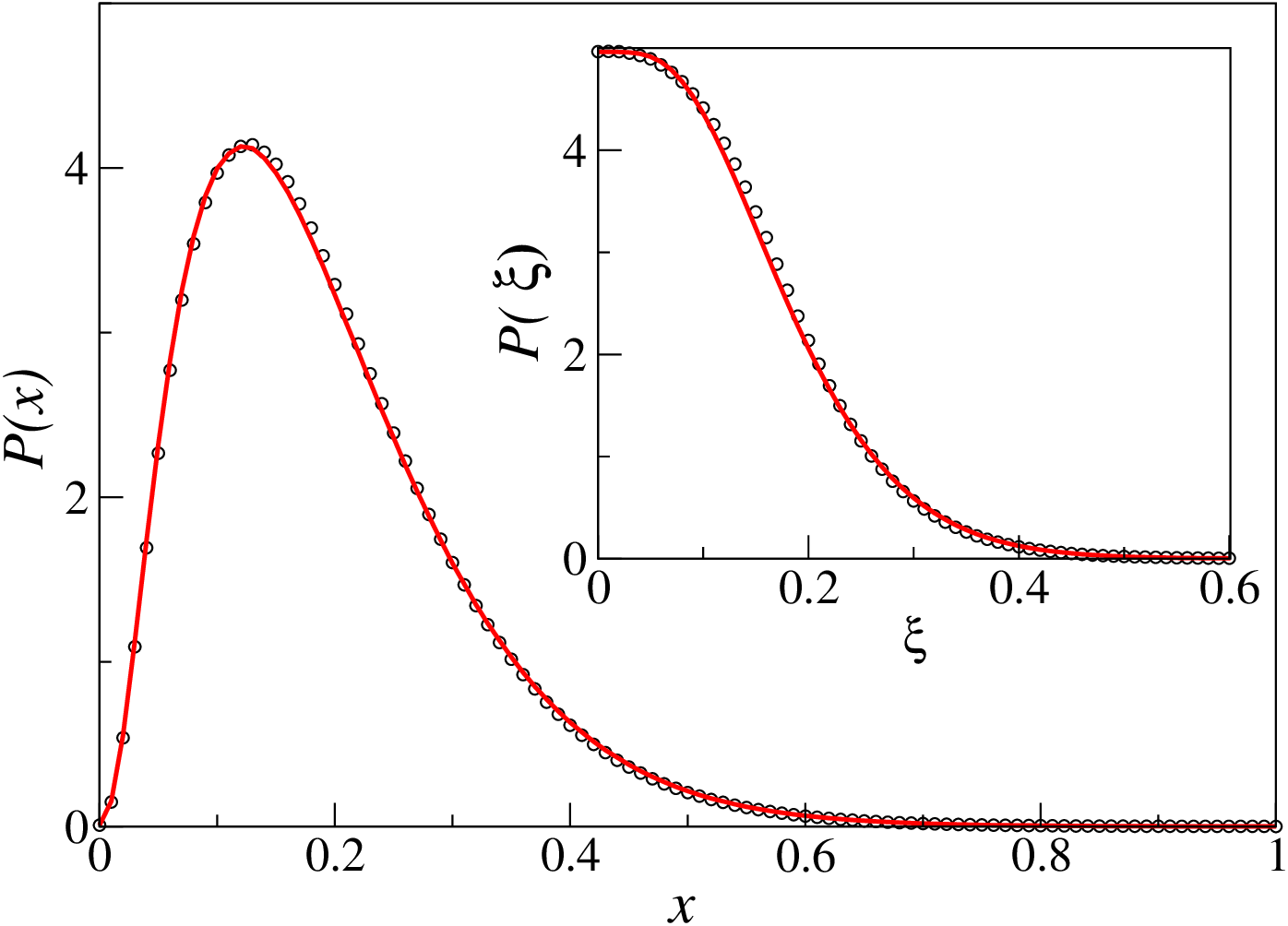}}
\caption{Comparison of wealth distribution of a tagged agent $k$, having 
$\lambda_k=0.4$ in CCM model (symbol) and the CC model (line) where 
savings propensity is identical ($\lambda=0.4$) for every agent. 
Average wealth of CC model  is taken as $w_k=0.198$, which is the average 
wealth of the tagged agent. For both CCM and CC model, $N=100$. 
In the inset, we compare the  distribution of "noise" in these systems.}
\label{fig:CC-CCM} 
\end{figure}

To see  why the dynamics of CCM model is same as Eq. (\ref{eq:model}) 
we look at the  steady state  wealth distribution of a  tagged agent $k$ 
whose savings propensity is $\lambda_k=\lambda$. Wealth of 
this agent, which fluctuates in time about the mean  $w_k \sim 1/(1-\lambda)$,  
is distributed  as $P^{CCM}_\lambda(x)$.  
Now we take a system (namely the CC model) of  $N$ agents  with total 
wealth $w_kN$ and all agents having the same savings propensity as that 
of  the tagged agent in CCM model,  $i.e, \{\lambda_i=\lambda\}$. 
Numerically calculated steady state distribution of wealth
$P^{CC}_\lambda(x)$
in this case is found  to be identical to   $P^{CCM}_\lambda(x)$. 
As an  example, we take a tagged agent in CCM having $\lambda=0.4$  and   
compare  $P^{CC}_\lambda(x)$  and  $P^{CCM}_\lambda(x)$  
in Fig. \ref{fig:CC-CCM}. An excellent agreement suggests that agents in CCM 
model are independent, unaware of  savings propensity of other agents.
Since each agent in  CCM model trade with every other agent, 
it is not surprising that this model turns out to be mean-field system of 
non-interacting agents (it is also observed in \cite{Manna}),  
which  suggests and supports that one can replace (\ref{eq:cc}) by a 
single equation 
\be
x_i^\prime = \lambda x_i + \eta,  
\label{eq:single}
\ee
where $\eta= r T_{ij}$ is the noise. Note that  $x_j$ satisfies 
the same equation because PDF of $r$ is same as that of $1-r$. 
Replacement of the conserving dynamics (\ref{eq:cc})  by a single 
equation (\ref{eq:single}) which do not have conservation suggests that 
{\em conservation is not important} in these systems.

\begin{figure}
\vspace*{.2 cm}
\centerline{\includegraphics[width=6cm]{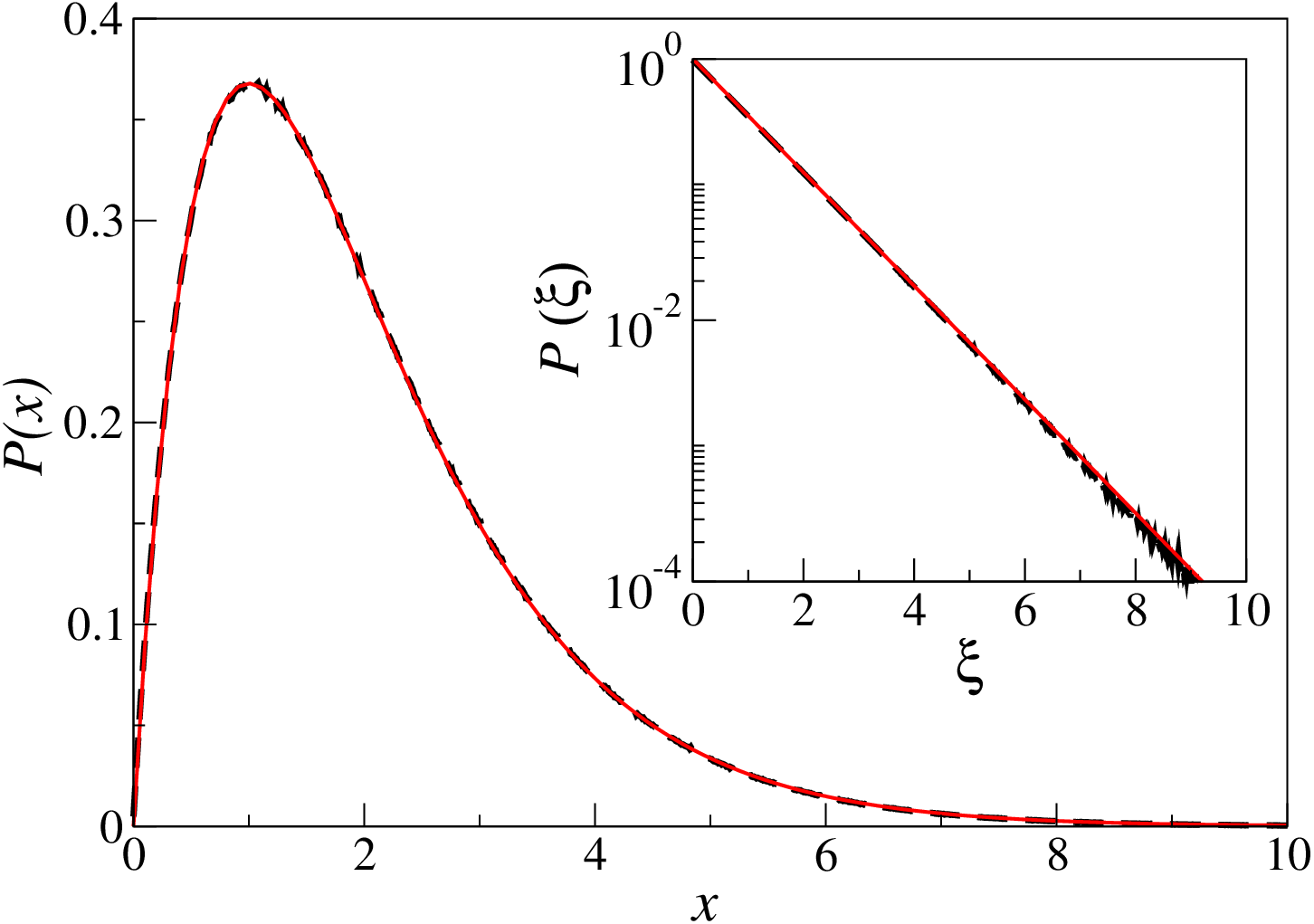}}
\caption{Wealth distribution $P(x)$ of a generic kinetic model 
(\ref{eq:new}) (dashed line), obtained from simulations (with $N=100$ and 
$\langle x \rangle=2$), is  compared with 
the steady state distribution $P(x)= x \exp(-x)$ (solid line) of 
corresponding non-conserving model (\ref{eq:non-cons}).
In the inset we compare PDF of noise $\xi$ for conserving (dashed line) and 
non-conserving (solid-line) models. For the later one $P(\xi) = \exp(-\xi)$.  }   
\label{fig:TwoStoch}
\end{figure}

  To emphasize this  point further that  'the wealth conserving dynamics can  
be replaced by a non-conserving one  similar to (\ref{eq:model})',   
we consider  other kinetic models. In a generic wealth conserving 
dynamics a pair of agents interact as follows, 
\bea
x_i^\prime &=& \lambda  x_i + \eta x_j \cr
x_j^\prime &=& (1-\lambda)  x_i + (1-\eta) x_j, 
\label{eq:new}
\eea
where both $\eta$ and $\lambda$ are stochastic variables with 
PDF ${\cal U}(x)$. We  will
prove, by mapping  wealth conserving dynamics (\ref{eq:new}) to a 
non-conserving one, that the steady state distribution of  this model is in 
fact $P(x)= x \exp(-x)\equiv \Gamma_2(x)$ (here $\langle x\rangle=2$). 
The non-conserving dynamics for this model is then  
\be
x^\prime = \lambda  x + \xi,
\label{eq:non-cons}
\ee 
where the noise  $\xi =  \eta \tilde x$ and $\tilde x$ is the wealth of 
the other agent in the conserving dynamics. 
Both $x$ and $\tilde x$ have the same distribution in the steady state.
If that distribution is $\Gamma_2(x)$, the PDF of   
$\xi$  is $P(\xi) = \exp(-\xi)$\cite{gamma*uniform}. 
Thus,  the dynamics of this model  is effectively the same as 
that of  the annealed $\lambda$ case of the AR model with 
exponential noise studied earlier in this article,
where the  steady state distribution is $P(x)= \Gamma_2(x)$.   
We have done numerical simulation of the 
conserving dynamics (\ref{eq:new}) and calculated the distribution of 
$\xi= \eta \tilde x$, and the steady state distribution of wealth $P(x)$. 
The resulting distributions are found (see Fig. \ref{fig:TwoStoch}) to be 
$P(\xi) = \exp(-\xi)$ and $P(x)= x \exp(-x)$, as expected from the 
non-conserving dynamics. Finally, we take the special case of 
the model with $\eta = \lambda$, which is the kinetic model 
studied in \cite{Yako00}, where the steady state distribution is 
$P(x) = \exp(-x)$. One may write a non-conserving dynamics in this case 
as 
\be
x^\prime = \lambda ( x + \tilde x).
\label{eq:yako}
\ee  
Again, both $x$ and $\tilde x$ have the same distribution in steady state. 
If $P(x)= exp(-x)$, then  using \cite{gamma*uniform,exp+exp} one can show that 
PDF of right hand side of (\ref{eq:yako}) is same as that of the left hand side.
These  generic examples thus strongly suggest that  both conserving and 
non-conserving dynamics  approaches the same steady state. 

In conclusion,  we introduce a simple model  which captures 
the growing feature of realistic markets.  Agents in these models do 
not involve in direct trading, they interact only through  the market.
Their net gain  depend on how much they invest and how much they gain 
from the market. The market, naturally noisy,  is modeled by a stochastic 
variable  having a specific distribution with fixed or 
varying mean. In our model,  return from the market  is considered to be 
independent of individual agent's investment (a natural extension  would be 
when $\xi(t)$ explicitly depends on $x(t)$). One of our main results is 
that, the average wealth
of agents generically follow Pareto-distribution. 
%The detailed nature 
%of distribution depends only on (a) the  mean (or instantaneous mean in 
%case of growing markets) and (b) distribution of investment capacity of agents.  
We also argue that, when   average wealth of the market grows 
adiabatically, the wealth distribution of agents at any  given time
depends only on the instantaneous  market. For $static$ markets, 
exact steady state wealth distribution  of agents was calculated 
for a few cases. More importantly, the dynamics of usual  wealth 
conserving kinetic models studied  in econophysics  as  
model markets   can be mapped to the dynamics of our model which 
does not have conservation.

Auto regression is a usual  technique for economists, for study of 
financial time series. These new models,  being auto-regressive 
in nature, build  connections between standard auto-regressive models and 
other kinetic models of markets.  Kinetic models which  are believed 
to be the only model explaining Pareto-law for the tail of the wealth 
distribution is not quite correct. In particular both,  conservation  
of wealth during each trading and global conservation of wealth are 
not necessary for obtaining Pareto-distribution. Auto-regression is 
an alternative which is more generic.
 
%%%%%%%%%%%%%%%%%%%%%%%%%%%%%%%%%%%%%%%%%%%%%%%%%%%%%%%%
Acknowledgments : One of the authors (UB) would like to acknowledge thankfully the 
financial support of the Council of Scientific and Industrial Research, 
India (SPM-07/489(0034)/2007)
%%%%%%%%%%%%%%%%%%%%%%%%%%%%%%%%%%%%%%%%%%%%%%%%%%%%%%


\begin{thebibliography}{99}
\bibitem{gamma*uniform}    If PDF of $x$ and $y$ are $\Gamma_2(x)$ and 
${\cal U}(y)$ respectively then  PDF of $u=xy$ is $\exp(-u)$.
\bibitem{exp+exp}   If PDF of $x$ and $y$ are $\exp(-x)$ and 
$\exp(-y)$ respectively then  PDF of $u=x+y$ is $u\exp(-u)$.
\end{thebibliography}
\end{document}